\documentclass[12pt,showkeys,onecolumn,showpacs,preprintnumbers,amsmath,amssymb]{revtex4}

\usepackage{graphicx}
\usepackage{dcolumn}
\usepackage{bm}

\begin{document}

\preprint{Chaos/xxx}

\title{Synchronizabilities of Networks: A New index \\}
\author{Huijie Yang$^1$}
\email{huijieyangn@eyou.com}
\altaffiliation {Corresponding author}
\author{Fangcui Zhao$^{2,3}$}
\author{Binghong Wang$^1$}
\address{$^1$ Department of Modern Physics,
         University of Science and Technology of China,Anhui Hefei 230026, China\\
         $^2$ College of Life Science and Bioengineering, Beijing University of Technology,
         Beijing 100022, China\\
         $^3$ Institute of Theoretical Physics, Beijing University of
         Technology, Beijing 100022, China}
\date{\today}

\begin{abstract}
The random matrix theory is used to bridge the network structures
and the dynamical processes defined on them. We propose a possible
dynamical mechanism for the enhancement effect of network
structures on synchronization processes, based upon which a
dynamic-based index of the synchronizability is introduced in the
present paper.
\end{abstract}


 \maketitle \textbf{The impact of network structures on
the synchronizability of the identical oscillators defined on them
is an important topic both for theory and potential applications.
From the view point of collective motions, the synchronization
state is a special elastic wave occurring on the network, while
the initial state is a abruptly assigned elastic state. The
synchronizability should be the transition probability between the
two states. By means of the analogy between the collective state
and the motion of an electron walking on the network, we can use
the quantum motion of the electron to find the motion
characteristics of the collective states. The random matrix theory
(RMT) tells us that the nearest neighbor level spacing
distribution of the quantum system can capture the dynamical
behaviors of the quantum system and the corresponding classical
system. A Poison distribution shows that the transition can occur
only between successive eigenstates, while a Wigner distribution
shows that the transition can occur between any two eigenstates. A
Brody distribution, an intermediate between the two extreme
conditions, can give us a quantitative description of the
transition probability. Hence, it can be used as an index to
represent the synchronizability. As examples, the Watts-Strogatz
(WS) small-world networks and the Barabasi-Albert(BA) scale-free
networks are considered in this paper. Comparison with the widely
used eigenratio index shows that this index can describe the
synchronizability very well. It is a dynamic-based index and can
be employed as a measure of the structures of complex networks.}

\section{Introduction}
 Recent years witness an avalanche investigation of complex
networks \cite{1,2,3}. Complex systems in diverse fields can be
described with networks, the elements as nodes and the relations
between these elements as edges. The structure-induced features of
dynamical systems on networks attract special attentions, to cite
examples, the synchronization of coupled oscillators
\cite{4,5,6,7,8}, the epidemic spreading \cite{9,10} and the
response of networks to external stimuli \cite{11}.

 Synchronization is a wide-ranging phenomenon which can be found in
social, physical and biological systems. Recent works show that
some structure features of complex networks, such as the
small-world effect and the scale-free property, can enhance
effectively the synchronizabilities of identical oscillators on
the networks, i.e., synchronization can occur in a much more wide
range of the coupling strength.

We consider a network of $N$ coupled identical oscillators
\cite{12}. The network structure can be represented with the
adjacent matrix $A$, whose element $A_{ij} $ is $0$ and $1$ if the
nodes $i$ and $j$ are disconnected and connected, respectively.
Denoting the state of the oscillator on the node $i$ as $x^i =
\left( {{\begin{array}{*{20}c}
 {x_i } \hfill \\
 {p_i } \hfill \\
\end{array} }} \right)$,the
dynamical process of the system is governed by the following
equations,

\begin{equation}
\label{eq1} \dot {x}^i = F(x^i) - \sigma \sum\limits_{j = 1}^N
{L_{ij} Q(x^j)} ,
\end{equation}

\noindent where $\dot {x}^i=F(x^i)$ governs the individual motion
of the $i$th oscillator, $\sigma$ the coupling strength and
$Q(x^j)$ the output function. The matrix $L$ is a Laplacian
matrix, which reads,

\begin{equation}
\label{eq2}
\begin{array}{l}
 L_{ij} = \left\{ {{\begin{array}{*{20}c}
 {  \sum\limits_s^N {A_{is} } =  k_i } \hfill & {(i = j)} \hfill \\
 {-A_{ij} } \hfill & {(i \ne j)} \hfill \\
\end{array} }} \right. \\
 {\begin{array}{*{20}c}
 \hfill & { =  k_i \delta _{ij} - A_{ij} } \hfill \\
\end{array} }, \\
 \end{array}
\end{equation}

\noindent where $k_i$ is the degree of the node $i$, i.e., the
number of the nodes connecting directly with the node $i$. The
eigenvalues of $L$ are real and nonnegative and the smallest one
is zero. That is, we can rank all the possible eigenvalues of this
matrix as $0=\gamma _1 \le \gamma _2 \le \cdots \le \gamma _{N} $.
Herein, we consider the fully synchronized state, i.e., $\left|
{x^i(t) -x^j(t)} \right| \to 0$ as $t\to \infty$ for any pair of
nodes $i$ and $j$.

  Synchronizability of the considered network of oscillators can be
quantified through the eigenvalue spectrum of the Laplacian matrix
$L$ . Here we review briefly the general framework established in
\cite{12,13}. The linear stability of the synchronized state is
determined by the corresponding variational equations, the
diagonalized $N$ block form of which reads, $\dot {z} = \left[
{DF(s) + \gamma DQ(s)} \right]z$. $z$ is the different modes of
perturbation from the synchronized state. For the $i$th block, we
have $\gamma=\sigma \gamma_i$,$i=1,2,\cdots,N$. The synchronized
state is stable if the Lyapunov exponents for these equations
satisfy $\Gamma(\sigma \gamma_i)<0$ for $i=2,3,\cdots,N$. Detailed
investigations \cite{12,13} show that for many dynamical systems,
there is a single interval of the coupling strength
$(\sigma_1,\sigma_2)$, in which all the Lyapunov exponents are
negative. In this case, the synchronized state is linearly stable
if and only if
$\frac{\gamma_N}{\gamma_2}<\frac{\sigma_2}{\sigma_1} \equiv \chi$.
While $\chi$ depends on the the dynamics, the eigenratio
$R=\frac{\gamma_N}{\gamma_2}$ depends only on the topological
structure of the network. Hence, this eigenratio represents the
impacts of the network structure on the networks's
synchronizability. This framework has stimulated an avalanche
investigation on the synchronization processes on complex
networks. It has been widely accepted as the quantity index of the
synchronizability of networks.

However, the eigenratio is a Lyapunov exponent-based index. It can
guarantee the linear stability of the synchronized state. It can
not provide enough information on how the network structure
impacts the dynamical process from an arbitrary initial state to
the final synchronized state. How the structures of complex
networks impact the synchronization is still a basic problem to be
understood in detail. In this paper, by means of the random matrix
theory (RMT) , we try to present a possible dynamical mechanism of
the enhancement effect, based upon which we suggested a new
dynamic-based index of the synchronizabilities of networks.

\section{Dynamic-based Index of Synachronizability}
 The RMT was developed by Wigner, Dyson, Mehta,
and others to understand the energy levels of complex quantum
systems, especially heavy nuclei \cite{14}. Because of the
complexity of the interactions, we can postulate that the elements
of the Hamiltonian describing a heavy nucleus are random variables
drawn from a probability distribution and these elements are
independent with each other. A series of remarkable predictions
are found to be in agreement with the experimental data. The great
successes of RMT in analyzing complex nuclear spectra has
stimulated a widely extension of this theory to several other
fields, such as the quantum chaos, the time series analysis
\cite{15,16,17,18,19}, the transport in disordered mesoscopic
systems, the complex networks \cite{20,21,22,23,24,25,26,27,28},
and even the QCD in field theory. For the complex quantum systems,
the predictions represent an average over all possible
interactions. The deviations from the universal predictions are
the clues that can be used to identify system specific, non-random
properties of the system under consideration.

One of the most important concepts in RMT is the nearest neighbor
level spacing (NNLS) distribution \cite{14}. Enormous experimental
and numerical evidence tells us that if the classical motion of a
dynamical system is regular, the NNLS distribution of the
corresponding quantum system behaves according to a Poisson
distribution. If the classical motion is chaotic, the NNLS
distribution will behave in accordance with the Wigner--Dyson
ensembles, i.e, $\sim s \cdot \exp ( - \kappa s^2 ) $. $s$ is the
NNLS. The NNLS distribution of a quantum system can tell us the
dynamical properties of the corresponding classical system. This
fact is used in this paper to bridge the structure of a network
with the dynamical characteristics of the dynamical system defined
on it.

From the state of the considered system, $X = \left(
{{\begin{array}{*{20}c}
 {x_i } \hfill \\
 {p_i } \hfill \\
\end{array} }} \right)\left| {i = 1,2, \cdots ,N} \right.$, we can
construct the collective motion of the system as,

\begin{equation}
\label{eq3} \Phi (i,t)= x_i(\theta_i,I_i,t,Y),i=1,2,\cdots,N,
\end{equation}
\noindent where $\theta_i$ and $I_i$ are the phase and the
amplitude of the oscillator $i$. $Y$ is the other
oscillation-related parameters. $\Phi$ describes the elastic wave
on the considered network and $Eq(3)$ presents the displacements
at the positions $i=1,2,\cdots,N$ at time $t$. Because of the
identification of the oscillators, the individual motions should
behave same except the phases and the amplitudes. The
synchronization process can be described as the transition from an
arbitrary initial collective state, $ \left(
{{\begin{array}{*{20}c}
 {\theta _1^0 } \hfill & {\theta _2^0 } \hfill & \cdots \hfill & {\theta
_N^0 } \hfill \\
 {I_1^0 } \hfill & {I_2^0 } \hfill & \cdots \hfill & {I_N^0 } \hfill \\
\end{array} }} \right)
$, to the final fully synchronized state, $ \left(
{{\begin{array}{*{20}c}
 {\theta _1^{syn} } \hfill \\
 {I_1^{syn} } \hfill \\
\end{array} }} \right) = \left( {{\begin{array}{*{20}c}
 {\theta _2^{syn} } \hfill \\
 {I_2^{syn} } \hfill \\
\end{array} }} \right) = \cdots= \left( {{\begin{array}{*{20}c}
 {\theta _N^{syn} } \hfill \\
 {I_N^{syn} } \hfill \\
\end{array} }} \right)
$.

\noindent The probability of the transition should be the
synchronizability of the considered network. The larger the
transition probability, the easier for the system to achieve the
fully synchronized state.

The collective states are the elastic waves on the considered
network. This kind of classical waves are analogous with the
quantum wave of a tight-binding electron walking on the network.
They obey exactly a same wave equation. In literature\cite{29,30,31,32},
this analogy is used to extend the
concept of Anderson localization state to the classical phenomena
as elastic and optical waves. In this paper we will use it to find
a quantitative description of the transition probability between
the collective states.

The tight-binding Hamiltonian of an electron walking on the
network reads,

\begin{equation}
\label{eq4} H = \sum\limits_{i = 1}^N {\varepsilon _i \cdot \left|
i \right\rangle \left\langle i \right|} + \sum\limits_{i \ne j}^N
A_{ij}{t_{ij} \cdot \left| i \right\rangle \left\langle j \right|}
,
\end{equation}

\noindent where $\varepsilon_i$ is the site energy of the $i$th
oscillator, $t_{ij}$ the hopping integral between the nodes $i$
and $j$. Because of the identification of the oscillators, all the
site energies are same, denoted with $\varepsilon_i=\varepsilon$.
Generally, we can set $\varepsilon=0$ and $t_{ij}=1$, which leads
to the relation $H=A$. Ranking the spectrum of $A$ as $\lambda _1
\ge \lambda _2 \ge \cdots \ge \lambda _N $, we denote the
corresponding quantum states with $\Psi _i \left| {i = 1,2,3,
\cdots ,N} \right.$. Hence, the NNLS distribution of the adjacent
matrix $A$ can show us the dynamical characteristics of the
collective motions.

If the NNLS obeys the Poisson form, the transition probability
between two eigenstates $\Psi _m $ and $\Psi _n $ will decrease
rapidly with the increase of $\left| {\lambda _m \left. { -
\lambda _n } \right|} \right.$, and the transition occurs mainly
between the nearest neighboring eigenstates. This state is called
quantum regular state. If the NNLS obeys Wigner form, the
transitions between all the states in the same chaotic regime the
initial state belongs to can occur with almost same probabilities.
The electron is in a quantum chaotic state.

The corresponding collective states of the classical dynamical
system to the quantum chaotic and regular sates are called
collective chaotic and collective regular states, respectively. If
the dynamical system is in a collective chaotic state, the
collective motion modes in same chaotic regimes can transition
between each other abruptly, while if the system is in a
collective regular state only the neighboring collective motion
modes can transition between each other. Generally, a dynamical
system may be in an intermediate state between the regular and the
chaotic states, which is called soft chaotic state.

The NNLS distribution can be obtained by means of a standard
procedure. The first step is the so-called unfolding. In the
theoretical predictions for the NNLS, the spacings are expressed
in units of average eigenvalue spacing. Generally, the average
eigenvalue spacing changes from one part of the eigenvalue
spectrum to the next. We must convert the original eigenvalues to
new variables, called unfolded eigenvalues, to ensure that the
spacings between adjacent eigenvalues are expressed in units of
local mean eigenvalue spacing, and thus facilitates comparison
with analytical results. Define the cumulative density function
as, $G(\lambda_m)=N \int_{-\infty}^{\lambda_m} g(\lambda) d
\lambda$, where $g(\lambda)$ is the density of the original
spectrum. Dividing $G(\lambda)$ into the smooth term
$G_{av}(\lambda)$ and the fluctuation term $G_f(\lambda)$, i.e.,
$G(\lambda)=G_{av}(\lambda)+G_f(\lambda)$, the unfolded energy
levels can be obtained as,

\begin{equation}
\label{eq5} \xi_m=G_{av}(\lambda_m).
\end{equation}

If the system is in a soft chaotic state, the NNLS distribution
can be described with the Brody form \cite{33}, which reads,

\begin{equation}
\label{eq6} P(s) = \frac{\beta }{\eta } \cdot s^{\beta - 1} \cdot
\exp \left[ { - \left( {\frac{s}{\eta }} \right)^\beta } \right].
\end{equation}

\noindent We can define the accumulative probability distribution
as, $Q(s) = \int_{ - \infty }^s {P(u)du} $. The parameter $\beta $
can be obtained from the linear relation as follows,

\begin{equation}
\label{eq7} lnT(s) \equiv ln\left[ {ln\left( {\frac{1}{1 - Q(s)}}
\right)} \right] = \beta lns - \beta ln\eta .
\end{equation}

\noindent For the special condition $\beta = 1$, the probability
distribution function (PDF) $P(s)$ degenerates to the Poisson form
and the system is in a regular state. For another condition $\beta
= 2$, the PDF obeys the Wigner-Dyson distribution $P(s) \propto s
\cdot \exp ( - \gamma \cdot s^2)$ and the system is in a hard
chaotic state. If the system is in an intermediate soft chaotic
state, we have, $1 < \beta < 2$.

Hence, from the perspective of random matrix theory, the
synchronizability can be described with the parameter $\beta $.
The larger the value of $\beta$, the easier for the system to
become fully synchronized. By this way we find a possible
dynamical mechanism for the enhancement effects of the network
structures on the synchronization processes.

\section{Results}
In reference \cite{27}, the authors prove that the spectra of the
Erdos-Renyi, the Watts-Strogatz(WS) small-world, and the growing
random networks (GRN) can be described in a unified way with the
Brody distribution. Herein, we are interested in the relation
between the parameter $\beta $ and the eigenratio $R$. Detailed
works show that $R$ is a good measure of the synchronizability of
complex networks, especially the small world and scale-free
networks \cite{34,35,36,37}.

Figure 1 shows the relation between $\beta $ and $R$ for WS
small-world networks  \cite{38}. We use the one-dimensional
regular lattice-based model. In the regular lattice each node is
connected with its $d$ right-handed neighbors. Connecting the
starting and the end of the lattice,  with the rewiring
probability $p_r $rewire the end of each edge to a randomly
selected node. In this rewiring procedure self-edge and double
edges are forbidden. Numerical results for WS small-world networks
with $N = 3000$ and $d = 2$ are presented. We can find that the
Brody distribution can capture the characteristics of the PDFs of
the NNLS very well, as shown in the panel (a) in Fig.1. With the
increase of $p_r $, the parameter $\beta $ increases rapidly from
$1.32$ to $\sim 1.85$, while the parameter $R$ decreases rapidly
from $3000$ to $\sim 10^1$. Hence, there exists a monotonous
relation between the two parameters $\beta$ and $R$.

Figure 2 gives the results for Barabasi-Albert (BA) scale-free
networks \cite{39}. Starting from a seed of several connected
nodes, at each time step connect a new node to the existing graph
with $w$ edges. The preferential probability to create an edge
between the new node and an existing node $f$ is proportional to
its degree, i.e., $p_{BA} \propto k(f)$. Numerical results for BA
scale-free networks with $N = 3000$ and $w = 1,2,3,\cdots,10$ are
presented. All the PDFs of the NNLS obey the Brody distribution
almost exactly. With the increase of $w$, the parameter $\beta $
increases from $0.35$ to $\sim 1.70$, while the parameter $R$
decreases from $51509$ to $\sim 10^1$. We can find also a
monotonous relation between the two parameters $\beta$ and $R$.

  For $w=1$, we have $\beta=0.35<1$. That is, rather than the
"repulsions" or un-correlations between the levels, there are a
certain "attractiveness" between the levels. In the construction
of the BA networks with $w=1$, each time only one node is added to
the existing network. The resulting network is a tree-like
structure without loops at all. Dividing the network into
subnetworks, we can find that many of them have similar
structures, which leads their corresponding level-structures being
almost same. Because of the weak coupling between the subnetworks,
the total level structure can be produced just by put all the
corresponding levels together. This kind of level-structure will
lead many NNLS tending to zero. Hence, $\beta<1$ is an extreme
case induced by tree-like structure. This special kind of
tree-like BA networks can not enhance the synchronization at all.

\section{Discussions}
In summary, by means of the NNLS distribution we consider the
collective dynamics in the networks of coupling identical
oscillators. For the two kinds of networks, we can find the
monotonous relation between the two parameters $\beta $ and $R$.
This monotonous relation tells us that the high synchronizability
is accompanied with a high extent of collective chaos. The
collective chaos may increase significantly the transition
probability of the initial random state to the final synchronized
state. The collective chaotic processes may be the dynamical
mechanism for the enhancement impacts of network structures on the
synchronizabilities.

The parameter $\beta $ in the NNLS distribution can be a much more
informative measure of the synchronizability of complex networks.
It reveals the information of the dynamical processes from an
arbitrary initial state to the final synchronized state. It can be
regarded in a certain degree as the bridge between the structures
and the dynamics of complex networks.

  One paradox may be raised about the argument in the present paper.
The Wigner distribution implies a larger correlation between the
eigenstates of the network than does the Poisson distribution. At
the same time, one can reverse the argument that Wigner
distribution implies level repulsion and, therefore, different
frequencies of oscillation of the normal modes, and therefore no
synchronization when these modes are coupled. It should be
emphasized that the eigenratio $R$ and the index $\beta$ should be
used together to capture the impacts of the network structures on
the synchronization processes. $R$ represents the linear stability
of the synchronized state, but it can not tell us how the final
synchronized state is reached from the initial state. On the other
hand, $\beta$ provides us a possible mechanism for this dynamical
processes, but it can not tell us the transition orientation. $R$
and $\beta$ reflect some features of the impacts of the network
structures on the synchronization processes, but there may be some
new important features to be found.

\section{Acknowledgement}
This work was supported by the National Science Foundation of
China under Grant No.70571074, No.70471033 and No.10635040. It is
also supported by the Specialized Research Fund for the Doctoral
program of Higher Education (SRFD No. 20020358009). One of the
authors would like to thank Prof. Y. Zhuo and J. Gu in China
Institute of Atomic Energy for stimulating discussions.

\pagebreak

\begin{figure}
\scalebox{1}[1]{\includegraphics{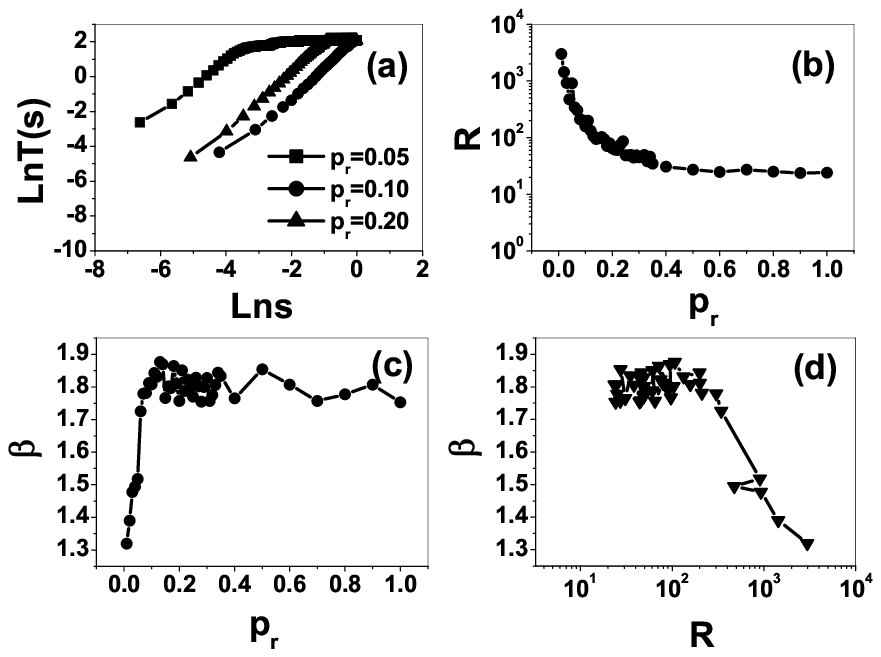}}
\caption{\label{fig:epsart}The relation of $\beta$ versus $R$ for
the constructed WS small-world networks. (a) Several typical
results for PDF of the NNLS. In the interested regions a brody
distribution can capture the characteristics very well. (b) With
the increase of the rewiring probability $p_r$ the eigenratio $R$
decreases rapidly. (c) With the increase of the rewiring
probability $p_r$ the parameter $\beta$ increases rapidly. (d) The
monotonous relation between the two parameters $\beta$ and $R$. }
\end{figure}

\begin{figure}
\scalebox{1}[1]{\includegraphics{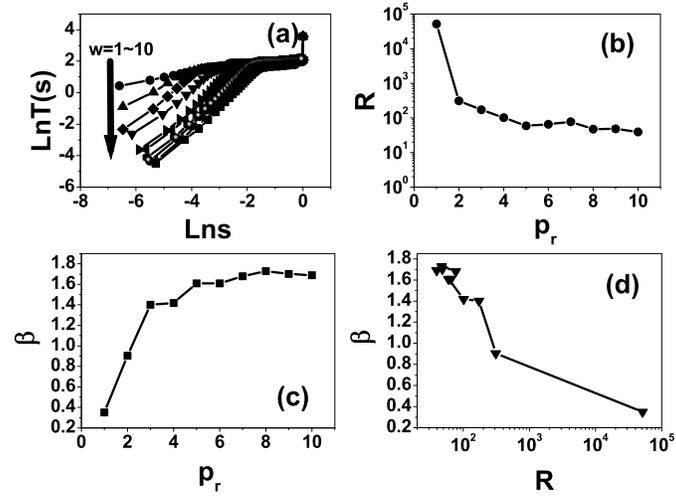}}
\caption{\label{fig:epsart}The relation of $\beta$ versus $R$ for
the constructed BA scale-free networks. (a) Results for PDF of the
NNLS. In the interested regions a brody distribution can capture
the characteristics very well. (b) With the increase of $w$ the
eigenratio $R$ decreases significantly. (c) With the increase of
$w$ the parameter $\beta$ increases significantly. (d) The
monotonous relation between the two parameters $\beta$ and $R$.  }
\end{figure}

\end{document}